\documentclass[aps,prl,twocolumn,groupedaddress]{revtex4}
\usepackage{graphics,epsfig}

\newcommand{\pdag}{{\phantom{\dagger}}}

\begin{document}


\title{Real Time Evolution in Quantum Many-Body Systems \\
With Unitary Perturbation Theory}


\author{A.~Hackl}
\affiliation{Institut f\"ur Theoretische Physik, Universit\"at zu K\"oln, Z\"ulpicher Str. 77, 50937 K\"oln, Germany}

\author{S. Kehrein}
\affiliation{Arnold Sommerfeld Center for Theoretical Physics and Center for Nanoscience (CeNS), Department f\"ur Physik,
Ludwig-Maximilians-Universit\"at M\"unchen}


\date{\today}

\begin{abstract}
We develop a new analytical method for solving real time evolution problems of quantum
many-body systems. Our approach is a direct generalization of the well-known canonical
perturbation theory for classical systems. Similar to canonical perturbation theory, secular
terms are avoided in a systematic expansion and one obtains stable long-time behavior.
These general ideas are illustrated by applying them to the spin-boson model and studying
its non-equilibrium spin dynamics.
\end{abstract}


\maketitle


The theoretical investigation of non-equilibrium quantum many-body systems
has recently become a very active field of research due to seminal experiments
in ultracold atomic gases (for example collapse and revival phenomena \cite{Greiner2002}),
electronic nanostructures (for example transport beyond the linear response regime \cite{Wiel2000}) 
and generally qubit dynamics in the presence of quantum dissipation. While non-equilibrium
classical systems have been long studied, quantum systems
in non-equilibrium hold the promise of many new phenomena yet to
be discovered. 
On the theoretical side, progress is hindered by the notorious difficulty
of solving non-equilibrium quantum many-body problems. Motivated by
the recent experiments, significant progress has been made with powerful
numerical methods like the time-dependent density matrix renormalization
group (TD-DMRG) \cite{Schollwoeck_TDDMRG} or the 
time-dependent numerical renormalization group (TD-NRG) \cite{Costi_nonequ,Anders_KM}. 
However, there are few reliable analytical methods available, especially
for non-perturbative problems (a noteable exeception is the real time 
RG method \cite{Schoeller}).

A key problem for analytical calculations is the appearance of {\it secular terms\/}
in time~$t$ that grow with some power of~$t$. Secular terms appear naturally
if one attempts a direct perturbative expansion, e.g.\ in the Heisenberg equations
of motion for the observables. Even if secular terms are multiplied by a small coupling
constant, they inevitably invalidate perturbation theory for large times even for
small coupling constants and make it impossible to draw conclusions about
the long-time behavior. This problem is also very well-known from classical
mechanics, dating back to studies of planetary motion in previous centuries.
In the context of analytical mechanics, its solution using {\it canonical 
perturbation theory\/} is well-established and
can be found in any textbook (see, for example, \cite{Textbook_analyticalmechanics}).
The basic idea is to first transform the Hamiltonian to normal form using
a canonical transformation. One can then easily solve the equations of motion
for the new position and conjugate momentum variables. Only after
integrating these equations of motion does one reexpress the old variables
in terms of the new time-evolved variables. It is well-established that this
yields a much improved long-time solution without any secular terms 
{\it even if the canonical transformation itself is only done perturbatively\/}.
Surprisingly, to the best of our knowledge to date no attempt has been made 
to implement an equivalent scheme based on unitary perturbation theory for 
quantum many-body systems. However, one key difference to classical systems
is that in quantum many-body systems one is often dealing with a continuous
energy spectrum, which makes naive unitary perturbation theory impossible
due to vanishing energy denominators. A way to solve this specific problem
has been established recently by means of the flow equation method
\cite{Wegner1994,Kehrein_STMP} (for related ideas see also the similarity 
renormalization scheme \cite{GlazekWilson}). The central idea of the 
flow equation method is to diagonalize a many-particle Hamiltonian through
a sequence of infinitesimal unitary transformations that eliminate interaction
matrix elements with large energy difference first before dealing with smaller 
energy differences. In this way one both reorganizes a perturbative expansion
in an RG-like manner, which allows one to recover non-perturbative energy
scales, and one avoids the above small energy denominator problem even
for a continuous energy spectrum.

In this Letter we develop the general framework for applying the flow
equation method to analytically solve real time evolution problems in 
quantum many-body systems in exact correspondence to canonical
perturbation theory in classical mechanics. We will see that likewise
secular terms are avoided and that one can obtain reliable results about
the long-time dynamics even in a perturbative framework. We will then
illustrate our approach by studying the real time evolution of the spin-boson
model with an initially polarized spin and a relaxed bath. 
The spin-boson model is the paradigm of dissipative quantum 
systems and its non-equilibrium behavior has recently been investigated using 
the TD-NRG method \cite{Anders_SB,Bulla_TDNRG_SB}, which motivates 
our choice. 

Let us briefly review the basic ideas of the flow equation approach
(for more details see \cite{Kehrein_STMP}). A many-body Hamiltonian~$H$
is diagonalized through a sequence of infinitesimal unitary transformations with
an anti-hermitean generator~$\eta(B)$,
\begin{equation}
\frac{dH(B)}{dB}=[\eta(B),H(B)] \ ,
\label{eqdHdB}
\end{equation}
with $H(B=0)$ the initial Hamiltonian.
The ``canonical" generator \cite{Wegner1994} is the commutator of the
diagonal part~$H_{0}$ with the interaction part~$H_{\rm int}$ of the Hamiltonian,
$\eta(B)\stackrel{\rm def}{=}[H_{0}(B),H_{\rm int}(B)]$. Under rather
general conditions the choice of the canonical generator leads to an increasingly
energy-diagonal Hamiltonian $H(B)$, where interaction matrix elements with
energy transfer~$\Delta E$ decay like $\exp(-B\,\Delta E^{2})$. For $B=\infty$
the Hamiltonian will be energy-diagonal and we denote parameters and operators in 
this basis by~$\tilde{~}$, e.g.\ $\tilde H=H(B=\infty)$.

The key problem of the flow equation approach is generically the generation of higher and higher
order interaction terms in (\ref{eqdHdB}), which makes it necessary to truncate
the scheme in some order of a suitable systematic expansion parameter (usually the
running coupling constant). Still, the infinitesimal nature of the approach makes it
possible to deal with a continuum of energy scales and to describe non-perturbative 
effects. This had led to numerous applications of the flow equation method
where one utilizes the fact that the Hilbert space is not truncated as opposed to
conventional scaling methods. Examples are the evaluation of correlation functions on all energy scales
in equilibrium problems \cite{Kehrein_STMP} and non-equilibrium problems, where one cannot
focus on low-energy degrees of freedom anyway (see, for example, the time-dependent
Kondo model \cite{LobaskinKehrein} or the Kondo model with voltage bias \cite{KehreinKMV}).

\begin{figure}
\includegraphics[clip=true,width=8.5cm]{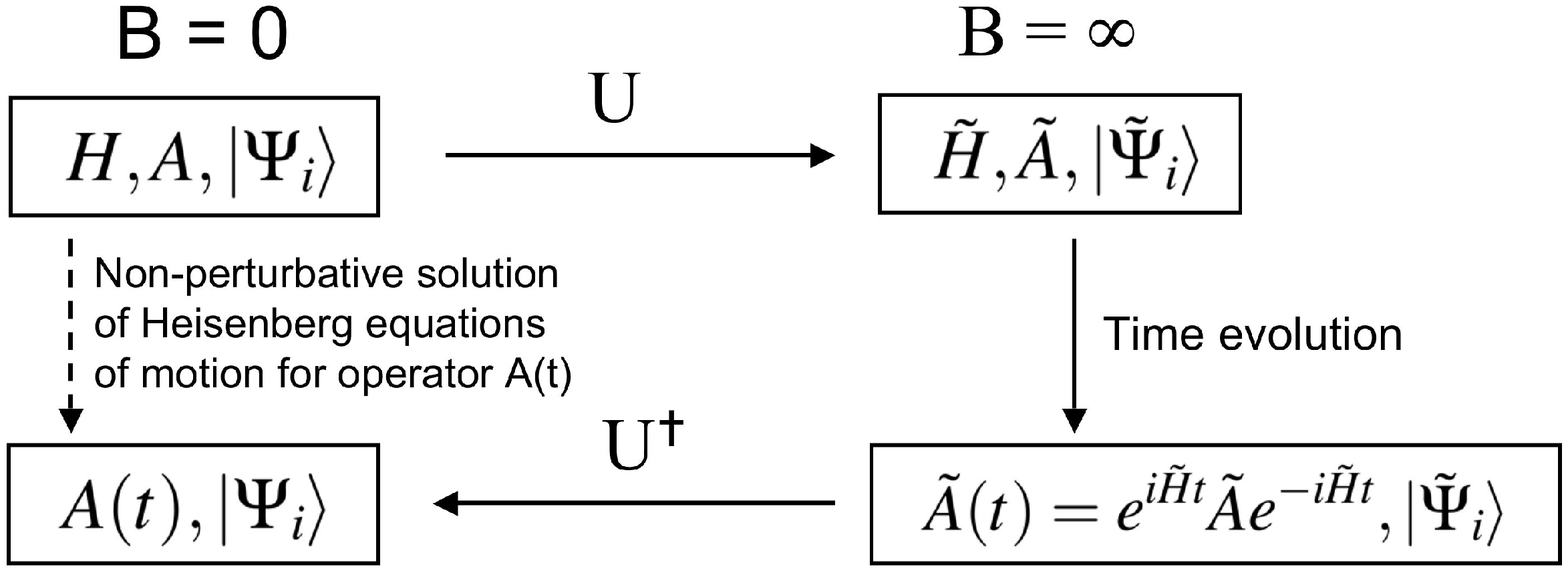}
\caption{\label{figrealtimefeq} The forward-backward transformation scheme induces
a non-perturbative solution of the Heisenberg equations of motion for an operator. 
$U$ denotes the full unitary transformation 
that relates the $B=0$ to the $B=\infty$ basis.\cite{footnotefullU}}
\end{figure}

We will now utilize these features to develop an analogue of canonical perturbation theory
in classical mechanics for quantum many-body problems. The general setup is described by
the diagram in Fig.~\ref{figrealtimefeq}, where $|\Psi_{i}\rangle$ is some initial
non-thermal state whose time evolution one is interested in. However, instead of following its
full time evolution it is more convenient to study the real time evolution of a given observable~$A$.
This is done by transforming the observable into the diagonal basis in Fig.~\ref{figrealtimefeq}
({\it forward transformation}):
\begin{equation}
\frac{dO(B)}{dB}=[\eta(B),O(B)] \ ,
\label{eqdOdB}
\end{equation}
with the initial condition $O(B=0)=A$. The central observation is that one can now solve the
real time evolution with respect to the energy-diagonal $\tilde H$
exactly, thereby avoiding any errors that grow proportional to time (i.e., secular terms):
this yields $\tilde A(t)$. Now
since the initial quantum state is given in the $B=0$~basis, one undoes the basis change by
integrating (\ref{eqdOdB}) from $B=\infty$ to $B=0$ ({\it backward transformation})
with the initial condition $O(B=\infty)=\tilde A(t)$.
One therefore effectively generates
a new non-perturbative scheme for solving the Heisenberg equations of motion for an
operator, $A(t)=e^{iHt}\,A(0)\,e^{-iHt}$, in exact analogy to canonical perturbation theory.
Notice that it is the last step of the backward transformation that distinguishes this scheme
from the flow equation evaluation of equilibrium correlation functions \cite{Kehrein_STMP}:
The equilibrium ground state or thermal states are in fact more easily expressed in the $B=\infty$
basis (since~$\tilde H$ is energy-diagonal) than in the $B=0$ (interacting) basis.
It should be mentioned that the same forward-backward transformation
scheme with respect to some given initial quantum state has recently also been successfully 
employed by Cazalilla \cite{Cazalilla2006}
for studying the nonequilibrium Luttinger model. The main difference to our approach is
that the bosonized Luttinger Hamiltonian becomes quadratic, which makes it possible
to work out the unitary transformation exactly in \cite{Cazalilla2006}
(the same holds in \cite{LobaskinKehrein}): therefore
stability questions regarding secular terms for a generic interacting system do not arise, 
which are the main focus of our work.

We now illustrate the general idea of our approach by studying the spin-boson model, which serves
as a paradigm in dissipative quantum physics and for qubit dynamics (for a review see, for example, 
\cite{Leggett}):
\begin{equation}
H=-\frac{\Delta}{2}\,\sigma_{x}+\frac{1}{2}\,\sigma_{z}\,\sum_{k} \lambda_{k} \,
(b^{\dagger}_{k}+b^{\pdag}_{k})+\sum_{k} \omega_{k}\,b^{\dagger}_{k}b^{\pdag}_{k} \ .
\label{eqdefSB}
\end{equation}
It describes a two state system coupled to a bath of harmonic oscillators. 
The effect of this dissipative environment is encoded in the spectral function
$J(\omega)\stackrel{\rm def}{=}\sum_{k}\lambda_{k}^{2}\,\delta(\omega-\omega^{\pdag}_{k})$.
In the sequel $\lambda_{k}$ is considered a small expansion parameter.
In this Letter we will only study the zero temperature case, $T=0$, although the generalization to
nonzero temperature is straightforward.
We use the following generator for the unitary flow \cite{KehreinMielke}:
\begin{eqnarray}
\eta(B)&=&i\,\sigma_{y}\sum_{k}\eta_k^{(y)} (b^{\pdag}_{k}+b^{\dagger}_{k})
+\sigma_{z}\sum_{k}\eta_k^{(z)} (b^{\pdag}_{k}-b^{\dagger}_{k}) \nonumber \\
&&+\sum_{k,l} \eta_{kl}\, :(b^{\pdag}_{k}+b^{\dagger}_{k})(b^{\pdag}_{l}-b^{\dagger}_{l}): \ ,
\end{eqnarray}
with $B$-dependent coefficients:
\begin{eqnarray}
\eta_{k}^{(y)}&=&-\frac{\lambda_{k}}{2}\,\Delta\,\frac{\omega_{k}-\Delta}{\omega_{k}+\Delta}
\ , \quad
\eta_{k}^{(z)}=-\frac{\lambda_{k}}{2}\,\omega_{k}\,\frac{\omega_{k}-\Delta}{\omega_{k}+\Delta}
\ , \nonumber \\
\eta_{kl}&=&\frac{\lambda_{k}\lambda_{l}\omega_{l}\Delta}{2(\omega_{k}^{2}-\omega_{l}^{2})}
\left(\frac{\omega_k-\Delta}{\omega_k+\Delta}  + \frac{\omega_l-\Delta}{\omega_l+\Delta}\right) \ .
\end{eqnarray}
Normal-ordering is denoted by $:\ldots :$, which serves as a systematic scheme to truncate the infinite sequence
of higher and higher operators generated by (\ref{eqdHdB}). Higher normal-ordered terms 
than the ones contained in (\ref{eqdefSB}) are 
neglected in the flow of the Hamiltonian, which amounts to neglecting small (of order~$\lambda_{k}^{2}$)
higher order cumulants in the Hamiltonian (this approximation is reliable for any super-Ohmic bath and for 
an Ohmic bath with $\alpha\lesssim 0.2$, for more details see 
\cite{KehreinMielke,Kehrein_STMP}). 
If one is interested in equilibrium properties, 
normal-ordering is performed
with respect to the equilibrium ground state, 
$b_{k}^{\pdag}b_{k'}^{\dagger}=:b_{k}^{\pdag}b_{k'}^{\dagger}:+\delta_{kk'}\,n(k)$,
where $n(k)$ is the Bose-Einstein distribution. However, later we will be interested in the real time evolution
of a non-thermal initial state~$|\Psi_{i}\rangle$. Hence, in order to minimize our truncation error,
we write more generally $b_{k}^{\pdag}b_{k'}^{\dagger}=:b_{k}^{\pdag}b_{k'}^{\dagger}:+
\delta_{kk'}\,n(k)+C_{kk'}$, where
$C_{kk'}\stackrel{\rm def}{=}
\langle\Psi_{i}\,|\,b_{k}^{\pdag}b_{k'}^{\dagger}\,|\,\Psi_{i}\rangle - \delta_{kk'}\,n(k)$.
The flow of $H(B)$ generated by this $\eta$ is 
\begin{eqnarray}
\frac{d\Delta}{dB}&=&-\Delta\,\sum_{k} \lambda_{k}^{2}\,
\frac{\omega_{k}-\Delta}{\omega_{k}+\Delta} 
\label{eqdDeltadB} \\
\frac{d\lambda_{k}}{dB}&=&-(\omega_{k}-\Delta)^{2}\,\lambda_{k} 
+2 \sum_{l} \eta_{kl}\,\lambda_{l}  \label{eqdlambdadB}  
\ .
\end{eqnarray}
The derivation of (\ref{eqdDeltadB}) and (\ref{eqdlambdadB}) is discussed in detail 
in~\cite{KehreinMielke,Kehrein_STMP}.
The diagonalized Hamiltonian for $B=\infty$ is
\begin{equation}
\tilde H=-\frac{\tilde\Delta}{2}\,\sigma_{x}+\sum_{k}\omega_{k}\,b^{\dagger}_{k} b^{\pdag}_{k} \ ,
\end{equation}
where $\tilde\Delta=\Delta(B=\infty)$ is the renormalized
tunneling matrix element. For example for an Ohmic bath, $J(\omega)=2\,\alpha\,\omega\,\Theta(\omega_{c}-\omega)$,
the renormalized tunneling matrix element derived from the solution of the flow equations~\cite{KehreinMielke,Kehrein_STMP} 
has the correct non-perturbative behavior \cite{Leggett}, 
$\tilde\Delta\propto\Delta\,\left(\Delta/\omega_{c}\right)^{\alpha/1-\alpha}$.

The observables in the $B=\infty$ basis are given by solving~(\ref{eqdOdB}) for a suitable
ansatz for the flowing observable \cite{Kehrein_STMP}.
For example,
\begin{eqnarray}
\sigma_{x}(B)&=&h(B)\,\sigma_{x}
+\sigma_{z}\sum_{k}\left(\chi_{k}(B)\,b^{\pdag}_{k}+\bar\chi_{k}(B)\,b^{\dagger}_{k}\right) \nonumber \\
&+&\alpha(B) 
+i\sigma_{y}\sum_{k}\left(\mu_{k}(B)\,b^{\pdag}_{k}-\bar\mu_{k}(B)\,b^{\dagger}_{k}\right)
\end{eqnarray}
where higher normal-ordered terms generated in~$O(\lambda_{k}^{2})$ during the flow (\ref{eqdOdB}) are again neglected.
The differential equations describing this flow take the following form:
\begin{eqnarray}
\frac{dh}{dB}&=&-\sum_{k} \left(\eta_{k}^{(y)}(\chi_{k}^{\pdag}+\bar\chi_{k}^{\pdag})
+\eta_{k}^{(z)}(\mu_{k}^{\pdag}+\bar\mu_{k}^{\pdag})\right) \nonumber \\
&&-4\sum_{k,l}\eta_{k}^{(y)}\,C_{kl}^{\pdag} (\chi_{l}^{\pdag}+\bar\chi_{l}^{\pdag}) \nonumber \\
\frac{d\chi_{k}}{dB}&=&2\,h\,\eta_{k}^{(y)}+\sum_{l} 
\left(\eta^{\pdag}_{kl}(\chi_{l}+\bar\chi_{l})+ \eta^{\pdag}_{lk}(\bar\chi_{l}-\chi_{l})\right) 
 \nonumber \\
\frac{d\mu_{k}}{dB}&=&2\,h\,\eta_{k}^{(z)}-\sum_{l} 
\left(\eta^{\pdag}_{lk}(\mu_{l}+\bar\mu_{l})+ \eta^{\pdag}_{kl}(\mu_{l}-\bar\mu_{l})\right) \nonumber \\
\frac{d\alpha}{dB}&=&\sum_{k} \left(\eta_{k}^{(y)}(\mu_{k}^{\pdag}+\bar\mu_{k}^{\pdag})
+\eta_{k}^{(z)}(\chi_{k}^{\pdag}+\bar\chi_{k}^{\pdag})\right) \ ,
\label{eqdsigmaxdB}
\end{eqnarray}
with the initial conditions $h(B=0)=1, \chi_{k}(B=0)=\mu_{k}(B=0)=\alpha(B=0)=0$.
For $\Delta\in {\rm supp}\: J(\omega)$ the observable decays completely, $\tilde h\stackrel{\rm def}{=} h(B=\infty)=0$, 
implying decoherence \cite{Kehrein_STMP}. The ground 
state expectation value of~$\sigma_{x}$ is then given by $\tilde\alpha\stackrel{\rm def}{=}\alpha(B=\infty)$.

\begin{figure}
\includegraphics[width=8.0cm]{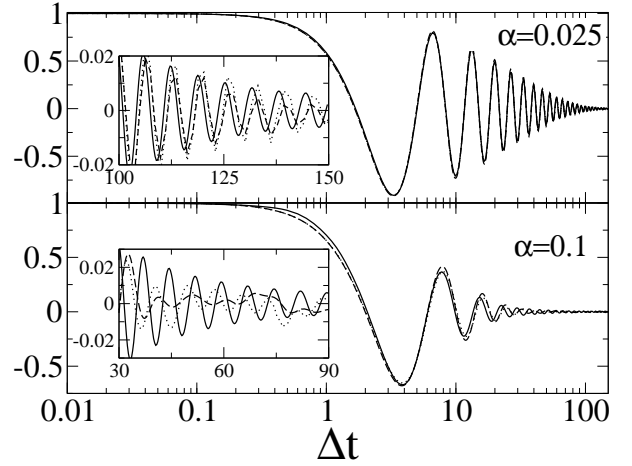}
\caption{\label{figrealtimeevol}
Real time evolution of the spin expectation value $\langle\sigma_{z}(t)\rangle$ starting from a polarized spin
in $z$-direction with a relaxed Ohmic bath (see text) for two different values of~$\alpha$ and 
$\omega_{c}/\Delta=10$. The full lines are the flow equation results, the dashed lines TD-NRG curves
for $\Lambda=2.0$ and the dotted lines for $\Lambda=1.41$. The TD-NRG results are courtesy of
F. Anders, see \cite{Anders_SB}. The various curves agree extremely well except for very long
times shown in the insets.}
\end{figure}

For real time evolution problems we now solve the Heisenberg equations of motion in the diagonal
basis, $\tilde\sigma_{z}(t)=e^{i\tilde H t} \tilde\sigma_{z} e^{-i\tilde H t}$. The result
is straightforward
\begin{eqnarray}
\tilde\chi_{k}(t)&=&\big(\tilde\chi_{k}(0)\cos(\tilde\Delta t)
+i\,\tilde\mu_{k}(0)\sin(\tilde\Delta t)\big)e^{-i\omega_{k} t} \\
\tilde\mu_{k}(t)&=&\big(\tilde\mu_{k}(0)\cos(\tilde\Delta t)
+i\,\tilde\chi_{k}(0)\sin(\tilde\Delta t)\big)e^{-i\omega_{k} t} \ , \nonumber 
\end{eqnarray}
while $\tilde h$ and $\tilde\alpha$ remain unchanged. In complete analogy to 
canonical perturbation theory, we next undo the unitary transformation (\ref{eqdOdB}).
The values of $\tilde h, \tilde\alpha, \tilde\chi_{k}(t), \tilde\mu_{k}(t)$ are used
as initial values in the system of differential equations (\ref{eqdsigmaxdB}) at $B=\infty$, which
is then integrated backwards to $B=0$. This yields $h(t), \alpha(t), \chi_{k}(t), \mu_{k}(t)$,
which parametrize the time-evolved operator~$\sigma_{x}$,
\begin{eqnarray}
\sigma_{x}(t)&=&h(t)\,\sigma_{x}
+\sigma_{z}\sum_{k}\left(\chi_{k}(t)\,b^{\pdag}_{k}+\bar\chi_{k}(t)\,b^{\dagger}_{k}\right) \nonumber \\
&+&\alpha(t) 
+i\sigma_{y}\sum_{k}\left(\mu_{k}(t)\,b^{\pdag}_{k}-\bar\mu_{k}(t)\,b^{\dagger}_{k}\right)
\end{eqnarray}
in the original basis of the problem. Thereby the forward-backward transformation scheme
induces a non-perturbative solution of the Heisenberg equations of motion, 
compare Fig.~\ref{figrealtimefeq}.

For the purposes of this Letter, we
focus on the numerical solution of the above differential equations 
by discretizing the bosonic bath with $O(10^{3})$ modes (notice that an approximate
analytical treatment is equally possible). 
The initial quantum state~$|\Psi_{i}\rangle$ is taken as spin up, 
$\langle\Psi_{i}|\sigma_{z}|\Psi_{i}\rangle=+1$, with a relaxed bath with respect to
this fixed spin. This yields $C_{kk'}=\lambda_{k}\lambda_{k'}/4\omega_{k}\omega_{k'}$.
We have implemented the
numerical solution for all components of the spin degree of freedom. In order to assess
the accuracy of our approach, 
the time evolution of $\langle \sigma_{z}(t)$ is shown in Fig.~\ref{figrealtimeevol}
and compared with TD-NRG data for two values of the discretization parameter~$\Lambda$.
One finds excellent agreement except for very long time scales (shown in the insets
of Fig.~\ref{figrealtimeevol}), where the TD-NRG discretization error becomes noticeable
(since the curves depend on~$\Lambda$). 
\begin{figure}
\includegraphics[clip=true,width=6.5cm]{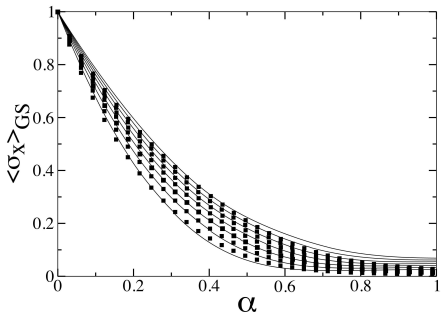}
\caption{\label{figcompNRG} Ground state expectation value of $\sigma_{x}$:
Comparison of flow equation results (curves) and NRG data (squares) from
\cite{compNRG} for an Ohmic bath with damping~$\alpha$. The
results are for  $\omega_{c}/\Delta=25, 28.6, 33.3, 40, 50, 66.7, 100$ 
from top to bottom.}
\end{figure}
The flow equation solution for the observable $\langle\sigma_x(t)\rangle$ shows that it
approaches its flow equation equilibrium expectation 
value~$\langle\sigma_{x}\rangle_{\rm GS} $ with
an absolute error below~$10^{-2}$  for long times.  
A comparison of~$\langle\sigma_{x}\rangle_{\rm GS} $ with exact numerical
results using NRG \cite{compNRG} in Fig.~\ref{figcompNRG} again shows very good agreement. 

Summing up, we have shown how to implement an analogous scheme to canonical 
perturbation theory for quantum many-body systems. Using a simple but non-trivial example, 
we could demonstrate that the well-established advantages of canonical perturbation theory
versus naive perturbation theory carry over to our unitary perturbation approach as well,
in particular the absence of secular terms in real time evolution problems. Our results
are stable in the long-time limit (see Figs.~\ref{figrealtimeevol},\ref{figcompNRG})
and can be improved systematically in a 
{\em uniform\/} manner (as a function of time) by 
higher orders of the calculation. The underlying scheme of infinitesimal unitary
transformations permits to study non-perturbative effects \cite{footnoteName}.
Similar to the role of canonical perturbation theory in analytical mechanics, our
approach should be useful for other real time evolution problems 
from impurity systems to lattice models in quantum many-body physics \cite{MoeckelKehrein}.

We thank F.~Anders for making the TD-NRG data in Fig.~2 available to us.
We acknowledge financial support through SFB~484 of the Deutsche Forschungsgemeinschaft,
the Center for Nanoscience (CeNS) Munich and the German Excellence Initiative via the 
Nanosystems Initiative Munich (NIM).


\begin{thebibliography}{11}

\bibitem{Greiner2002}
M.~Greiner, O.~Mandel, T.W. H\"ansch, and I. Bloch,
Nature {\bf 419}, 51 (2002).

\bibitem{Wiel2000}
W.G. van der Wiel {\it et al.\/}, Science 289, 2105 (2000).

\bibitem{Schollwoeck_TDDMRG}
U.~Schollw{\"o}ck and S.R.~White, 
in {\it Effective models for low-dimensional strongly correlated systems\/},
edited by G. Batrouni and D. Poilblanc (AIP, Melville, New York, 2006), p.~155.

\bibitem{Costi_nonequ}
T.A.~Costi,
Phys. Rev. B {\bf 55}, 3003 (1997).

\bibitem{Anders_KM}
F.B.~Anders and A.~Schiller, 
Phys. Rev. Lett. {\bf 95}, 196801 (2005).

\bibitem{Schoeller}
H. Schoeller, Lect. Notes Phys. {\bf 544}, 137 (2000).

\bibitem{Textbook_analyticalmechanics}
See, for example, 
H.~Goldstein, Ch.P.~Poole, and J.L.~Safko,
{\it Classical Mechanics\/}
(Addison-Wesley, Third edition, 2002).

\bibitem{Wegner1994}
F.~Wegner, Ann. Phys. (Leipzig) {\bf 3}, 77 (1994).

\bibitem{Kehrein_STMP}
S.~Kehrein, {\it The Flow Equation Approach to Many-Particle Systems\/},
(Springer, Berlin Heidelberg New York, 2006).

\bibitem{GlazekWilson}
S.D.~G{\l}azek and K.G.~Wilson, Phys. Rev. D {\bf 48}, 5863 (1993);
{\bf 49}, 4214 (1994).

\bibitem{Anders_SB}
F.B.~Anders and A.~Schiller,
Phys. Rev. B {\bf 74}, 245113 (2006).

\bibitem{Bulla_TDNRG_SB}
F.B.~Anders, R.~Bulla, and M.~Vojta,
Phys. Rev. Lett. {\bf 98}, 210402 (2007).

\bibitem{LobaskinKehrein}
D. Lobaskin and S. Kehrein, Phys. Rev. B {\bf 71}, 193303 (2005);
J. Stat. Phys. {\bf 123}, 301 (2006).

\bibitem{KehreinKMV}
S. Kehrein, Phys. Rev. Lett. {\bf 95}, 056602 (2005).

\bibitem{footnotefullU}
The full unitary transformation $U$ can be expressed as an $B$-ordered
exponential, $U=T_{B}\exp\left(\int_{0}^{\infty}\eta(B)\,dB\right)$.
However, this expression is only formally useful since it cannot
be evaluated without additional approximations.

\bibitem{Cazalilla2006}
M.A. Cazalilla, Phys. Rev. Lett. {\bf 97}, 156403 (2006).

\bibitem{Leggett}
A.J. Leggett {\it et al.}, Rev. Mov. Phys. {\bf 59}, 1 (1987).

\bibitem{KehreinMielke}
S. Kehrein and A. Mielke, Ann. Phys. (Leipzig) {\bf 6}, 90 (1997).

\bibitem{compNRG}
T.A.~Costi and R.H.~McKenzie, 
Phys. Rev. A {\bf 68}, 034301 (2003).

\bibitem{footnoteName}
A more accurate description of our method would therefore be "unitary renormalized perturbation theory".

\bibitem{MoeckelKehrein}
Since the submission of this work, the forward-backward scheme
has already been successfully used for studying an interaction quench in the Hubbard model:
M. Moeckel and S. Kehrein, Phys. Rev. Lett. {\bf 100}, 175702 (2008).

\end{thebibliography}
\end{document}